\documentclass[%
 prl,%
 cp,%
 amsmath,amssymb,%
 reprint,%
 groupedaddress,
]{revtex4-1}

\usepackage{graphicx}
\usepackage{dcolumn}
\usepackage{bm}
\usepackage{textcomp}
\usepackage{float}
\usepackage[mathlines]{lineno}

\begin{document}

\title[Spin-orbit torque induced electrical switching of antiferromagnetic MnN ]{Spin-orbit torque induced electrical switching of antiferromagnetic MnN}

\author{M. Dunz}
\author{T. Matalla-Wagner}
\author{M. Meinert}
\email{meinert@physik.uni-bielefeld.de}
\affiliation{ 
Center for Spinelectronic Materials and Devices, Department of Physics, Bielefeld University, D-33501 Bielefeld, Germany 
}

\date{\today}

\begin{abstract}
Electrical switching and readout of antiferromagnets allows to exploit the unique properties of antiferromagnetic materials in nanoscopic electronic devices. Here we report experiments on the spin-orbit torque induced electrical switching of a polycrystalline, metallic antiferromagnet with low anisotropy and high N\'eel temperature. We demonstrate the switching in a Ta / MnN / Pt trilayer system, deposited by (reactive) magnetron sputtering. The dependence of switching amplitude, efficiency, and relaxation are studied with respect to the MnN film thickness, sample temperature, and current density. Our findings are consistent with a thermal activation model and resemble to a large extent previous measurements on CuMnAs and Mn$_2$Au, which exhibit similar switching characteristics due to an intrinsic spin-orbit torque.
\end{abstract}

\maketitle
\section{Introduction}
The discovery of the electrical switching of antiferromagnetic CuMnAs via an intrinsic spin-orbit torque has triggered immense interest of researchers working in the field \cite{Wadley2016, Jungwirth2018}. Experiments verified the proposed switching mechanism via direct imaging and that the remarkable properties of antiferromagnets, such as insensitivity to external magnetic fields and terahertz dynamics, can be exploited in devices \citep{Grzybowski2017, Olejnik2017a, Olejnik2017b}. The so-called N\'eel-order spin-orbit torque (NSOT) has initially been predicted \cite{Zelezny2014} for another material, Mn$_2$Au, which is an antiferromagnet with a very high N\'eel temperature \cite{Barthem2013}. Several works verified that the NSOT is also present in this material \cite{Bodnar2018, Meinert2018, Zhou2018}. Recent studies by some of the authors of this article have demonstrated that thermal activation and thermal assistance via Joule heating are key features to the understanding and realization of stable multi-level devices made of Mn$_2$Au or CuMnAs \cite{Meinert2018, MatallaWagner2019}.

Only few metallic materials with suitable magnetic and crystallographic symmetry for the NSOT are known \cite{Watanabe2018}, which poses a significant challange for the development and integration of devices based on these materials. Very recent work demonstrated that spin-orbit torque induced switching of insulating epitaxial NiO layers via the spin Hall effect (SHE) of an adjacent Pt layer results in very similar switching characteristics \cite{Chen2018, Moriyama2018, Gray2018, Baldrati2018}. While the details of the underlying mechanism are under debate, it nevertheless opens a new route in antiferromagnetic spintronics. Similarly, it was shown that $\alpha$-Fe$_2$O$_3$ can be switched and that Mn$_2$Au can be manipulated via the SHE in a way distinct from the intrinsic NSOT \cite{Cheng2019, Zhang2019, Zhou2019}.

\begin{figure}[t]
\includegraphics[width=8.6cm]{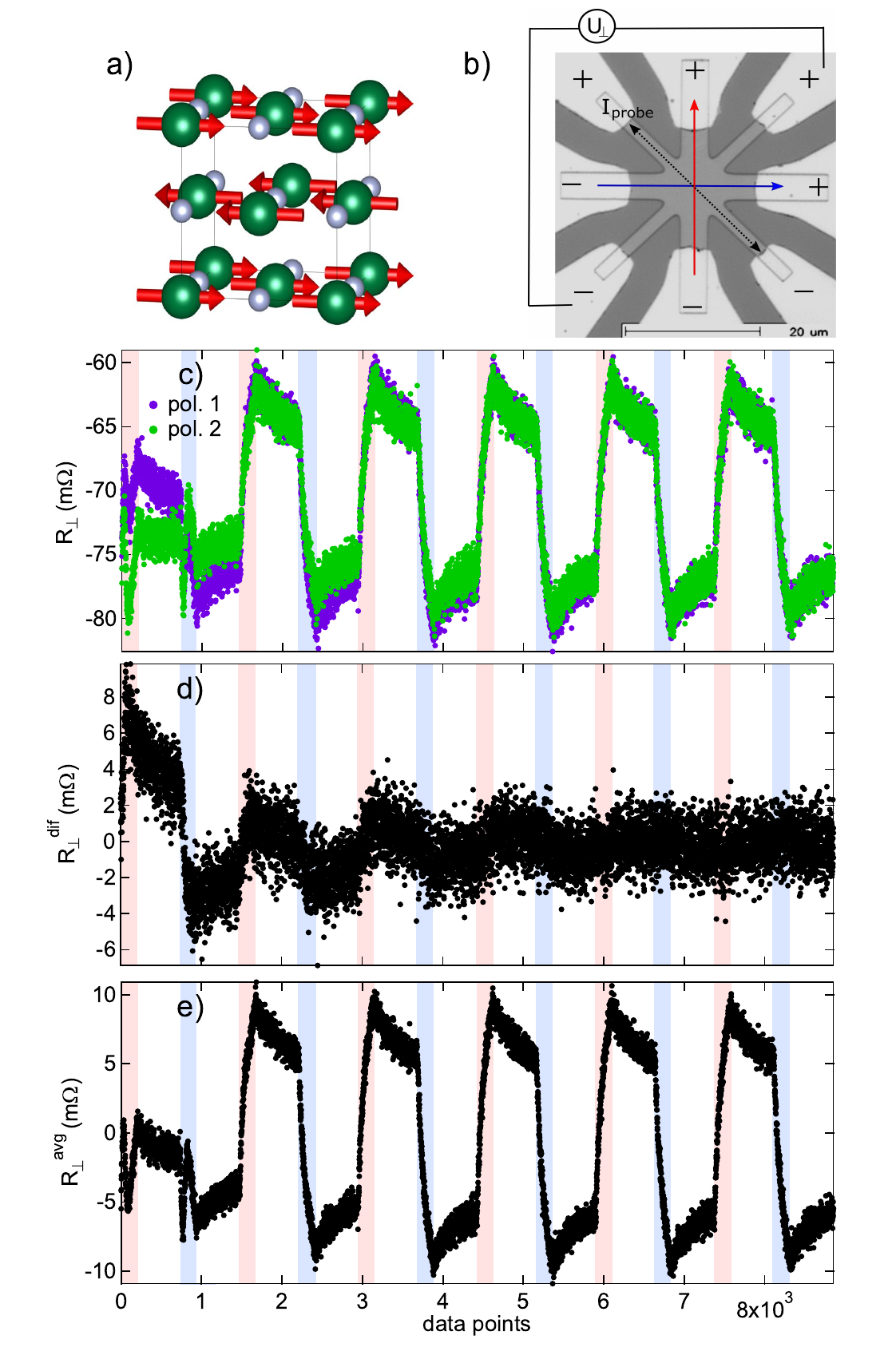}
\caption{\label{Fig1}a): Crystallographic and magnetic structure of MnN with AF-I N\'eel order. Larger, green balls represent Mn atoms, smaller grey balls represent N atoms. b) Micrograph of a star-shaped device used in the electrical experiment. The signs represent the connections to the positive/negative differential outputs and inputs of the voltage sources and the lock-in amplifier. The red and blue arrows represent the physical current directions of polarity 1. c) Raw $R_\perp$ traces from a switching experiment. Background colors correspond to pulsing the two pulse lines (red, blue) as sketched in b) or relaxation phase (white). During the pulse phase, the time between two bursts is approximately 2.2\,s, in the relaxation phase data are taken at 1\,s intervals. d) Difference between the two polarities in c). e) Average of the two polarities with the offset removed.}
\end{figure}

In the present article, we demonstrate that electrical switching is possible with polycrystalline, metallic antiferromagnets and an adjacent Pt layer. Thereby, we show that a much larger class of antiferromagnetic thin films can be manipulated via the SHE, including metallic and polycrystalline materials; the read-out is possible via either the planar Hall effect (PHE) \cite{Wadley2016, Meinert2018} or the spin Hall magnetoresistance (SMR) \cite{Chen2018, Moriyama2018}. In our experiment, we focus on a low-anisotropy antiferromagnet with high N\'eel temperature: MnN. It has a tetragonally distorted NaCl structure and a N\'eel temperature of 650\,K \cite{Suzuki2000, Leineweber2000}. Its magnetic structure is of the AF-I type with the magnetic moments aligned antiparallel along the (001) direction, see Fig. \ref{Fig1}\,a). However, the spin orientation is controversial and might depend critically on the lattice constants. In previous studies, some of the authors of this article have shown its utility for exchange bias applications with large exchange bias fields at room temperature \cite{Meinert2015, Zilske2017, Dunz2018a, Dunz2018b, Sinclair2019, Quarterman2019}. However, the critical thickness for the onset of exchange bias was observed to be around 10\,nm at room temperature, leading to the conclusion that MnN has a small magnetocrystalline anisotropy energy density \cite{Meinert2015}. Since the available torque from the SHE is not large, we decided to choose this low-anisotropy material, because it seems to be an ideal candidate for an electrical switching experiment. 

\section{Experiment}
We prepared Ta (6\,nm) / MnN ($t_{\text{MnN}}$) / Pt (4\,nm) samples on thermally oxidized Si substrates via dc magnetron sputter deposition at room temperature. The MnN layer was reactively sputtered from an elemental Mn target in a sputtering gas ratio of $50 \%$ Ar to $50 \%$ N$_2$, following the same procedure as reported in Ref. \onlinecite{Meinert2015}. Its (001) fiber-textured growth in the as-deposited state as descibed in detail in Ref. \onlinecite{Meinert2015} has been confirmed by x-ray diffraction. Magnetic and grain size characterization of similar films was performed previously using the so-called ``York protocol'' of exchange bias measurements and transmission electron microscopy \cite{OGrady2010, Sinclair2019}. In this study, the median lateral grain size of the MnN was found to be 4.8\,nm and the anisotropy constant at room temperature was estimated as $K_\mathrm{AF} \lessapprox 6 \times 10^5$\,J/m$^3$. In polarized neutron reflectometry measurements, the films were found to be slightly rich in nitrogen and no magnetic scattering from the MnN films could be detected \cite{Quarterman2019}. This excludes the possibility, that electrical switching of ferrimagnetic Mn$_4$N precipitates contributes to the signals we investigate in the present study. 

For the electrical switching experiments, the samples were patterned to star-shape \cite{Wadley2016} structures (cf. Fig. \ref{Fig1}b)) using electron beam lithography and Ar ion milling. The devices are connected to the measurement setup via Ta/Au contact pads and Au wire bonds. Our measurement system is identical to the one we used previously for the study of Mn$_2$Au and CuMnAs. It is described in detail in Ref. \onlinecite{MatallaWagner2019}. In all experiments presented here, we used a current pulse width of $\Delta t = 4\,$\mbox{\textmu s}. Pulses were grouped into bursts with constant charge per burst of $Q = 1.68 \times 10^{-4}$\,C and a duty cycle of 0.002. After every pulse, a delay of 2\,s was applied before taking the transverse resistance $R_\perp = U_\perp / I_\mathrm{probe}$ reading with a lock-in amplifier, cf. Fig. \ref{Fig1}\,b). To ensure constant nominal current density $j_0 = I_0 / (w d) = U_0 / (R w d)$, the pulse line resistances $R$ were measured before every switching cycle consisting of six repeats of 200 bursts per current direction and relaxation phases of 600\,s. Here, $j_0$ refers to a nominal current density with the total metallic film thickness $d$ and the current-line width $w = 4\,$\mbox{\textmu m}. To determine the current densities in the individual layers, we used a parallel conductor model to determine the Pt layer resistivity using the MnN and Ta resistivities of 180\,\textmu$\Omega$cm that were determined by four-point measurements on suitable reference samples. Additionally, the current density is corrected for the inhomogeneous current flow in the center-region of the star-structure by a factor of 0.6. For more details, we refer the reader to the Appendix to Ref. \onlinecite{MatallaWagner2019}. 

\section{Results}
In Fig. \ref{Fig1}\,c) we show typical raw data of switching with both polarities. To analyze a possible influence of the pulsing polarity, we calculate the differences and averages of the two polarities, see. Fig. \ref{Fig1} d) and e), respectively. While the first two repeats show a clear dependence on the polarity, further cycles show only negligible influence of the polarity. Due to the expected symmetry of the N\'eel order switching, we focus on the reproducible, polarity independent component of the measurement, i.e. the average over the two polarities after a training phase of three repeats. In the following, all switching traces refer to polarity-averaged switching traces after three repeats of training.

\begin{figure}[t]
\centering
\includegraphics[width=8.6cm]{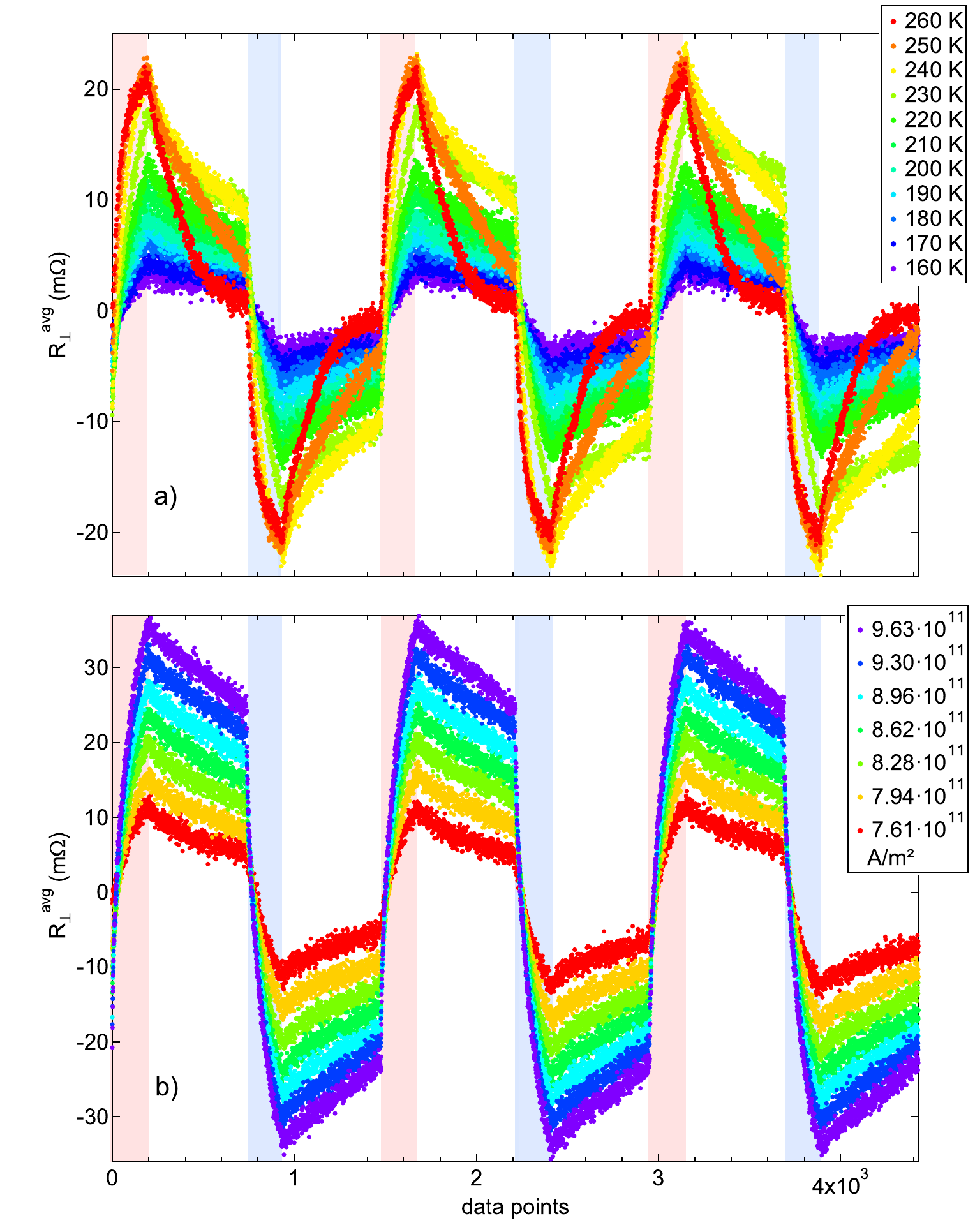}
\caption{\label{Fig2} a) Switching traces of the 6 nm MnN sample taken at temperatures between 140\,K and 260\,K. b) Switching traces of the same sample taken at 230 K for center-region Pt current densities of $j _\mathrm{Pt}= 7.61 \dots 9.63 \times 10^{11}$\,A/m$^2$.}
\end{figure}

In Fig. \ref{Fig2}, we show polarity-averaged switching traces for temperature dependence and current density dependence of the 6\,nm MnN sample. The temperature dependence shows clearly that higher temperature assists the switching process, with increasing steepness and amplitude. It is also clearly seen that with higher temperature, the relaxation becomes much faster and complete relaxation to the initial state is seen after 600\,s at 260\,K. The switching is also quite sensitive to the current density and large changes of the amplitude are seen within a fairly small interval of current densities. Recent work on the switching of insulating antiferromagnets with the SHE of Pt suggests that the typical ``saw-tooth'' shape of the transverse voltage traces is related to a degradation effect in the Pt film \cite{Cheng2019, Zhang2019}. To ensure that the electrical response in our experiment originates from the switching of the N\'eel order, we performed some reproducibility tests after cycling of the temperature. The degradation of the Pt layer should result in signals that are not reproducible after temperature- or current-cycling. Our results are, however, reproducible in the same device, which points to a magnetic origin.

To facilitate a quantitative analysis of the switching traces, we adopt the method from Ref. \onlinecite{MatallaWagner2019}. First, to remove the polarity-dependent component, we use the polarity-averaged datasets. Then we separate the switching traces into two regimes, namely pulsing along either the red or blue lines in Fig. \ref{Fig1}\,a) and relaxation. For the pulsing regime, we found a simple fit function consisting of a constant, an exponential function and a line appropriate. In this case, the variable is the burst count $b$:  
\begin{align}
\label{eq:pulsing_fit}
R_\text{p}(b) = c_0 + c_1 \, \exp \left( - \frac{b}{\mu} \right)  + c_2 \, b.
\end{align}
$c_{0,1,2}$ and $\mu$ are fitting parameters. Equation \ref{eq:pulsing_fit} is only a phenomenological fit function that helps to obtain the switching efficiency of the first burst $R_\mathrm{e}$ accurately by taking the derivative of Eq.~\eqref{eq:pulsing_fit} at $b=0$:
\begin{align}\label{eq:SwitchingEfficiency}
R_\text{e} = 
\left| 
\frac{d R_\text{p}(b)}{d b}
\right|_{b=0} 
= \left| -\frac{c_1}{\mu} + c_2 \right|.
\end{align}
For the relaxation regime, we use a simple exponential decay as a function of measurement time $t$ and add an offset:
\begin{align}\label{eq:relaxation_fit}
R_\text{r}(t) = d_0 + d_1 \, \exp \left( - \frac{t}{\tau_\mathrm{eff}}\right).
\end{align}
$d_{0,1}$ and $\tau_\mathrm{eff}$ are fitting parameters. The decay is characterized by the effective relaxation time constant $\tau_\mathrm{eff}$ for a given set of parameters. As we show in Ref. \onlinecite{MatallaWagner2019}, the exponential decay has a strict physical meaning. All antiferromagnetic grains of a polycrystalline film have volumes $V_\mathrm{g}$ which typically follow a lognormal distribution. These grains are related to the anisotropy energy barriers via $E_\mathrm{B} = K_\mathrm{AF} V_\mathrm{g}$. The relaxation time for the orientation of the N\'eel vector of a grain is given by the N\'eel-Arrhenius equation
\begin{equation}
\tau = f_0^{-1} \exp\left( \frac{E_\mathrm{B}}{k_\mathrm{B} T}  \right).
\end{equation}
Here $f_0 \approx 10^{12}\,\mathrm{s}^{-1}$ is the antiferromagnetic resonance frequency, $k_\mathrm{B}$ is the Boltzmann constant, and $T$ is the absolute temperature. During the pulsing, we excite grains with various energy barriers at the same time, where smaller $E_\mathrm{B}$ means that the N\'eel vector is easier to switch but will also relax faster. Therefore, one may expect to see a sum of multiple exponential decays during the relaxation phase. To simplify the analysis of the relaxation, we merge the ensemble of many different relaxation times into a single effective time constant $\tau_\mathrm{eff}$. The constant offset $d_0$ in the relaxation fits suggests that the switching has a long-term stable contribution. However, because of the time window of 600\,s for the observation of the relaxation, we can only tell there is a component in the signal that is stable for times substantially longer than this window. It is clear that $d_0$ is a function of the time window, so we refrain from a detailed analysis of this parameter. In addition, we define the difference of $R_\perp$ before and after applying the bursts along one current-line as the absolute switching amplitude $\left|\Delta R_\text{a}\right|$. 

\begin{figure}[t]
\centering
\includegraphics[width=8.6cm]{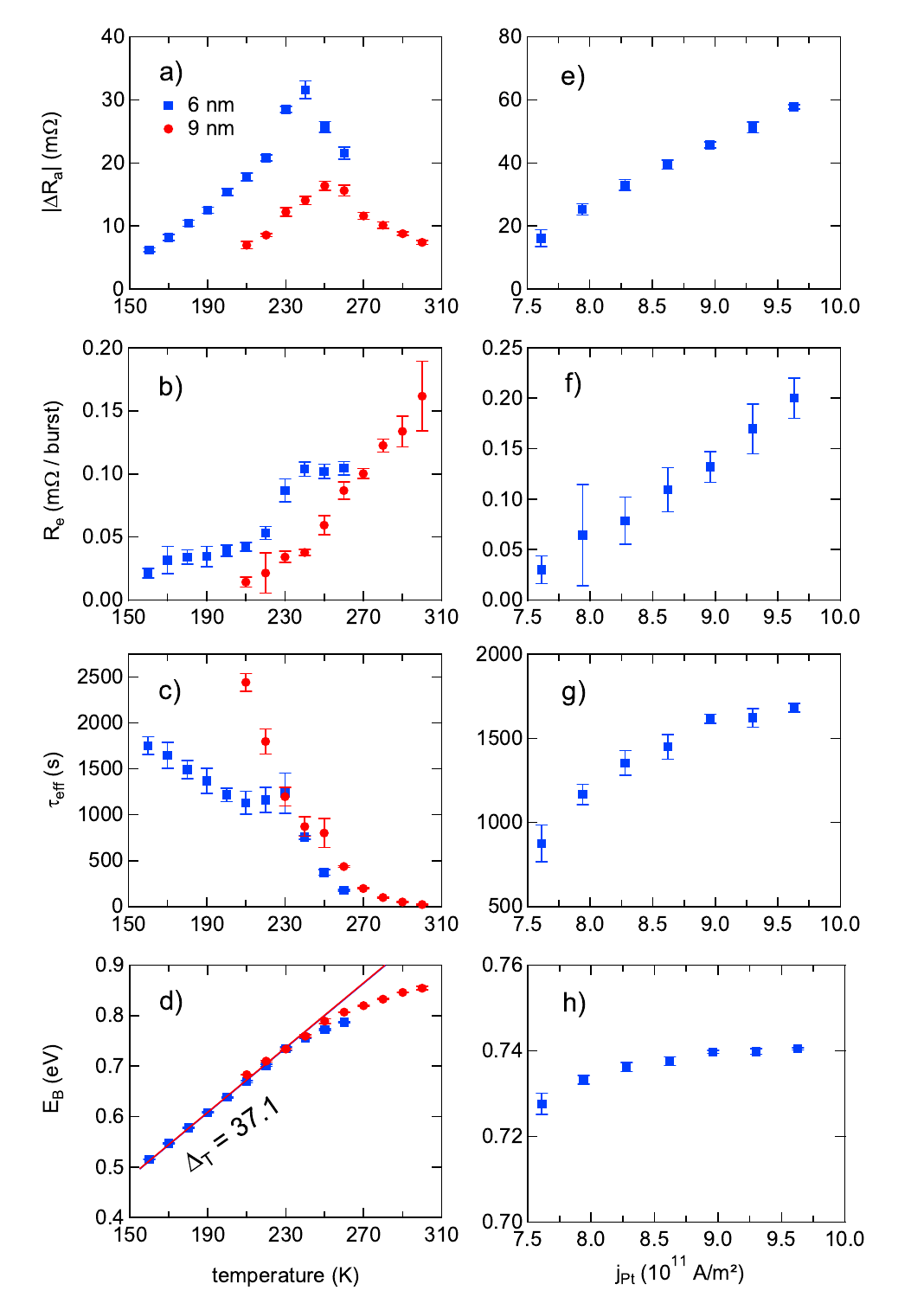}
\caption{\label{Fig3} a)-d) Temperature dependencies of a) absolute switching amplitude $\Delta R_\mathrm{a}$, b) switching efficiency $R_e$, relaxation time constant $\tau$, and switching barrier $E_\mathrm{B}$. e)-h) Pulse current density dependencies of d) $\Delta R_\mathrm{a}$ and $R_\mathrm{e}$,  f) relaxation time constant $\tau_\mathrm{eff}$, and h) switching barrier $E_\mathrm{B}$. The temperature dependence was measured with $j_0 = 5 \times 10^{11}$\,A/m$^2$, the current density dependence was measured at $T=230$\,K. Note that all Pt current densities are given for the center region of the star-structure.}
\end{figure}

In Fig. \ref{Fig3}, we summarize the results of this analysis for the temperature and current density dependencies. The absolute switching amplitude shows clear maxima for both film thicknesses as a function of the temperature (Fig. \ref{Fig3}\,a)). Remarkably, the thinner film shows a larger amplitude and the maximum is found at lower temperature. Simultaneously, the switching efficiency (Fig. \ref{Fig3}\,b)) increases with increasing temperature, but also shows an indication of peaking at a slightly higher temperature as compared to the amplitude. $\tau_\mathrm{eff}$ shows a very strong temperature dependence in both samples and is smaller for the 6\,nm MnN film thickness, see Fig. \ref{Fig3}\,c). This result is fully compatible with our thermal-activation model developed earlier for the switching in Mn$_2$Au and CuMnAs: the peak of the switching amplitude is loosely related to the maximum of the lognormal grain size distribution. The larger film thickness leads to a larger grain volume und thereby shifts the amplitude maximum to higher temperature. Simultaneously, higher temperatures lead to faster relaxation, as given by the N\'eel-Arrhenius equation. The energy barriers obtained from the relaxations are of the order $E_\mathrm{B} = 0.5 \dots 0.9\, \mathrm{eV}$, see Fig. \ref{Fig3}\, d). In contrast to naive expectation of scaling with film thickness, we find that the energy barriers are very similar in both films; we interpret this result later. As a function of current density, we find that both the switching amplitude (Fig. \ref{Fig3}\,e)) and the switching efficiency (Fig. \ref{Fig3}\,f)) are greatly increased with increasing current density. We find that $\tau_\text{eff}$ also depends on the current density (Fig. 3 g)) which is due to our simplification of taking a single effective realaxation time instead of observing the weights associated with many different relaxation time constants of the ensemble. At higher current density, the film temperature is substantially higher, which increases the proportion of larger grains with larger energy barriers that participate in the switching (Fig. 3h)). The observed $\tau_\text{eff}$ increases as these grains contribute to a slower relaxation at the measurement temperature. Notably, both films have effective relaxation time constants of less than 100\,s at room temperature; this is perfectly in line with the observation that exchange bias is observed with MnN only for larger film thicknesses of approximately 10\,nm \cite{Meinert2015} at room temperature. 

\begin{figure}[t]
\centering
\includegraphics[width=8.6cm]{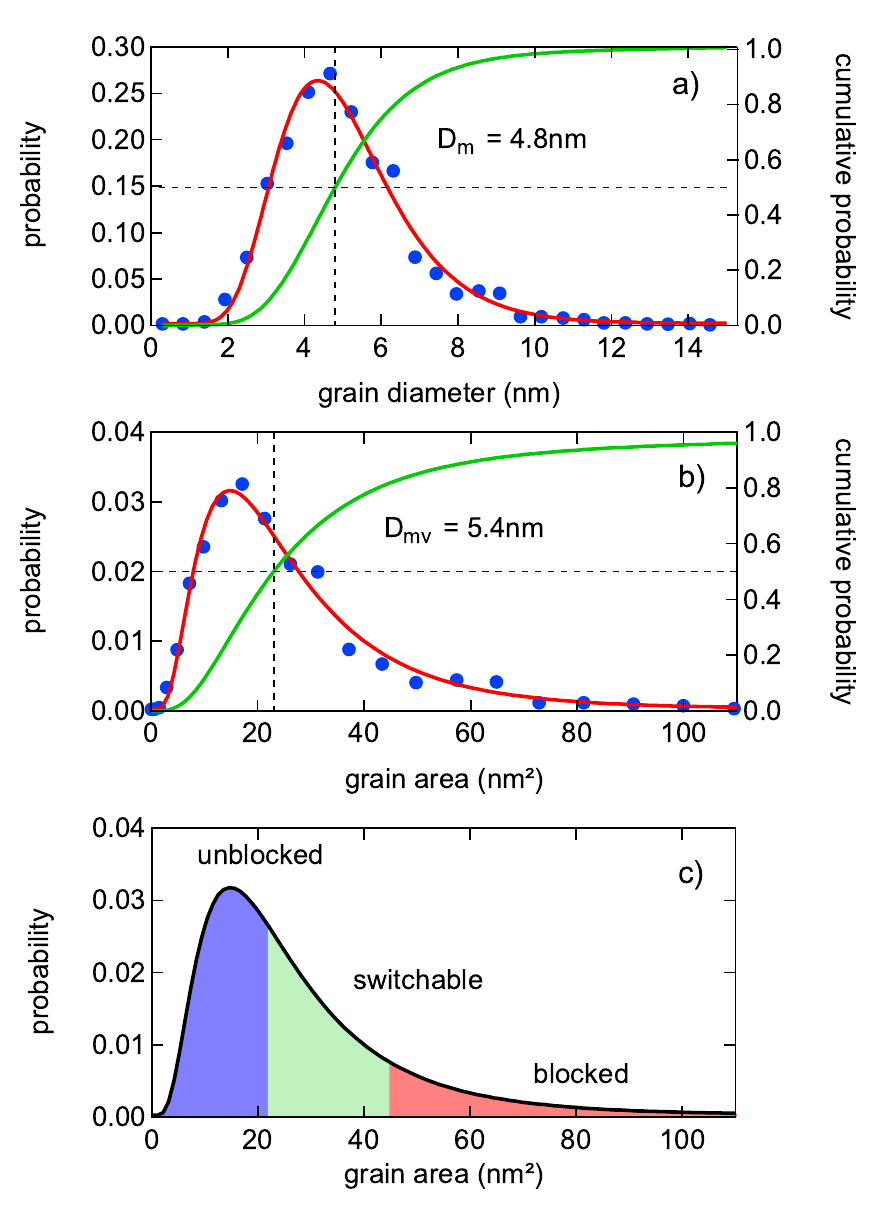}
\caption{\label{Fig4} a) Particle diameter analysis from a plan view TEM image. Data points taken from Ref. \onlinecite{Sinclair2019}. The integrated lognormal fit (i.e., the cumulative probability) indicates a median grain diameter of $D_\mathrm{m} = 4.8$\,nm. b) Particle size analysis recalculated for the grain area under the assumption of cylindrical grains. The lognormal fit indicates a median volume-derived grain diameter of $D_\mathrm{mv} = 5.4$\,nm. c) Model of the three grain-size regimes discussed in the main text. The three grain categories are drawn for the case of a MnN thickness of 9\,nm.}
\end{figure}

In Fig. \ref{Fig4}\,a), we reproduce the result of the particle diameter analysis from Ref. \onlinecite{Sinclair2019}. According to this analysis, the anisotropy energy density is $K_\mathrm{AF} \lessapprox 4 \times 10^5$\,J/m$^3$ for thin MnN films. For lognormal-distributed grain diameters, also the grain areas and grain volumes of cylindrical grains are lognormal distributed. The diameter of grains which correspond to the median volume is $D_\mathrm{mv} \approx 5.4$\,nm, see Fig. \ref{Fig4}\,b). This corresponds to $E_\mathrm{B} = 0.5$\,eV in the 9\,nm MnN film, which is clearly of the correct order of magnitude. However, this result indicates that the majority of the grains which contribute to the switching in our experiments are larger than the median of the distribution. The saturation of $E_\mathrm{B}$ as a function of temperature in Fig. \ref{Fig3}\,d) can thus be understood as a lack of grains with diameters larger than 7.5\,nm ($E_\mathrm{B} \approx 1$\,eV for the 9\,nm film). Indeed, according to the particle size analysis, less than 10\% of the grains have larger diameters. Therefore, their contribution to the electrical signal will be rather small. These results allow us to identify three classes of grains, which we call \textit{unblocked}, \textit{switchable}, and \textit{blocked}, see Fig. \ref{Fig4}\,c). The unblocked grains relax very quickly for all temperatures at which we performed measurements. The switchable grains correspond to the observed energy barriers of 0.5\dots0.9\,eV. Finally, the blocked grains remain blocked and are not switched with the spin-orbit torque. We note that only a narrow part of the switchable ensemble will contribute to the actual switching and relaxation at any given temperature. Correspondingly, one cannot directly relate the position of the switching amplitude maximum to the maximum of the grain size distribution, due to the complexity of the switching and relaxation dynamics: due to the Joule heating, the switching occurs at an elevated temperature, whereas the relaxation happens at the set measurement temperature. 

\begin{figure}[t]
\centering
\includegraphics[width=8.6cm]{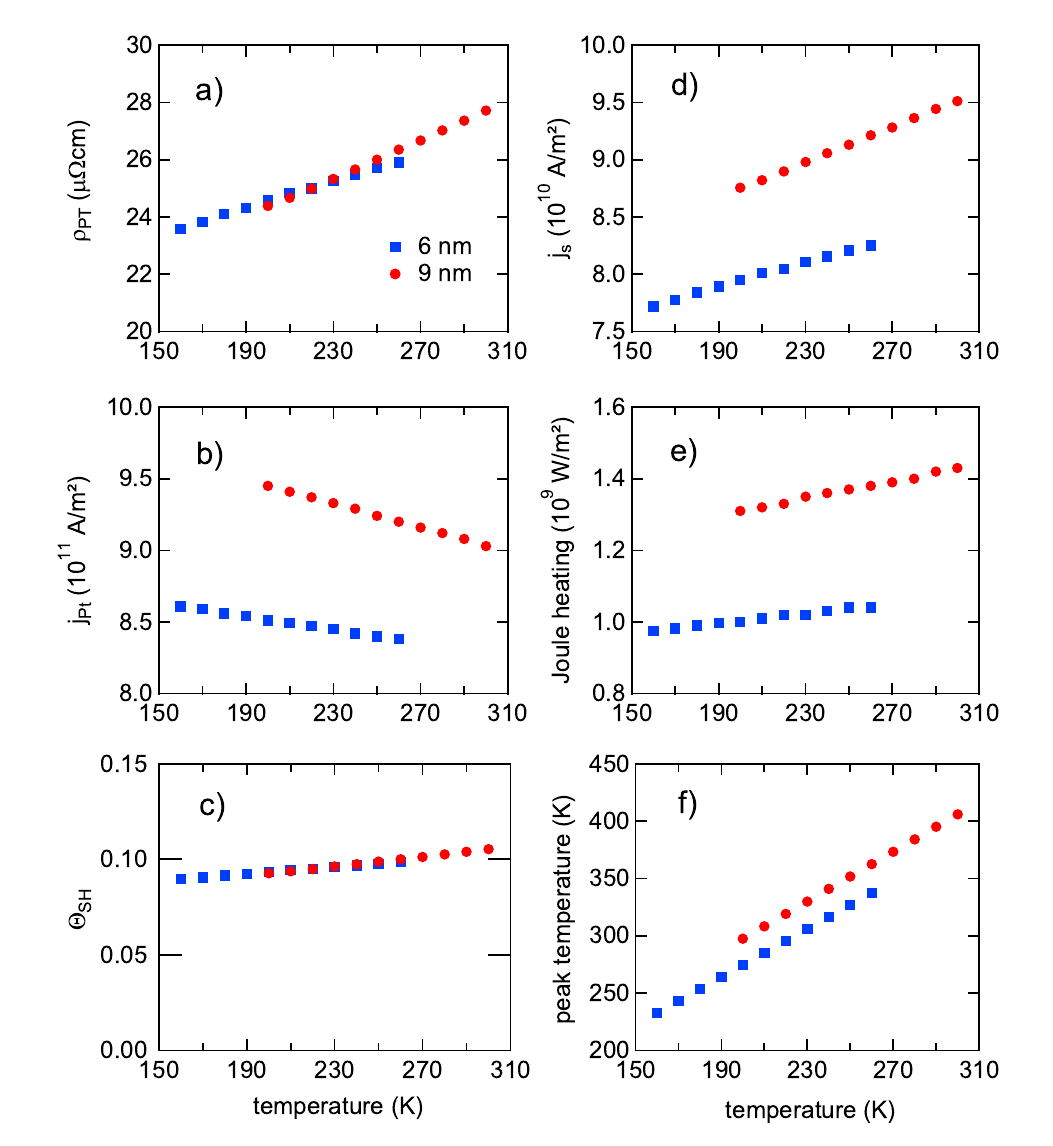}
\caption{\label{Fig5} Temperature dependencies of a) resistivity of the Pt layer, b) pulse current density in the center region of the Pt layer at $j_0 = 5 \times 10^{11}$\,A/m$^2$, c) calculated spin Hall angle, d) calculated spin current density, e) calculated Joule heating power, f) calculated peak temperature during the pulses. Note that all Pt current densities are given for the center region of the star-structure.}
\end{figure}

To shed further light on the Joule heating and the effect of the conducting multilayer system and associated shunting, we study the stack with the parallel resistor model and calculate the spin current density and Joule heating in the center-region as a function of the measurement temperature, see Fig. \ref{Fig5}. The model gives very similar Pt resistivities for the two samples (Fig. \ref{Fig5}\,a)), slight deviations probably arise from the neglect of the weak temperature dependence of the MnN and Ta resistivities. Because of the identical nomial current densities $j_0$, the sample with 9\,nm MnN has a larger center-region current density in the Pt layer, Fig. \ref{Fig5}\,b). On the basis of the resistivities and the center-region current densities, we calculate the spin Hall angles $\theta_\mathrm{SH} = \sigma_\mathrm{SH} \rho_\mathrm{Pt}$ (Fig. \ref{Fig5}\,c)) with $\sigma_\mathrm{SH} = 4 \times 10^{5}$\,$(\Omega \mathrm{m})^{-1}$ \cite{Obstbaum2016} and spin current densities $j_\mathrm{s} = \theta_\mathrm{SH} j_\mathrm{Pt}$ (Fig. \ref{Fig5}\,d)). Here, the spin current density from the Ta film is neglected, because the current density flowing in the Pt layer is approximately eight times larger. Because of the larger film thickness, the Joule heating power is larger in the 9\,nm MnN sample (Fig. \ref{Fig5}\,e)). Using the heating powers of the center-region of the star structures, we calculate the peak temperatures of the film using an analytical formula \cite{You2006, MatallaWagner2019}. A correction factor of 1.48 determined by a stationary finite-element simulation was applied to take the 50\,nm SiO$_2$ layer into account. Unsurprisingly, at identical measurement temperature, the film reaches a higher temperature due to the Joule heating with larger film thickness. Coming back to the similarity of $E_\mathrm{B}$ for different film thicknesses (Fig. \ref{Fig3}\,d)), we note that both thermal activation and the spin current density are larger in the thicker film. Both aspects lead to a more efficient switching in the thicker film, bringing the two thicknesses closer together in terms of efficiency. However, $E_\mathrm{B}$ is evaluated from the relaxation, which depends only weakly on how the state has been set, cf. Fig. \ref{Fig3}\,h). Both temperature dependencies of $E_\mathrm{B}$ in Fig. \ref{Fig3}\,d) can be fitted with identical line fits $E_\mathrm{B} = \Delta_T k_\mathrm{B} T$ with $\Delta_T = 37.1$ in the range up to 240\,K, while saturation is seen at higher temperature. We interpret this as a grain-selection process by the available torque. Only grains with $\Delta_T \approx 27 \dots 44$ can be switched and be observed to relax \cite{MatallaWagner2019}. Since the available torque is similar for all film thicknesses, in the thicker film grains with smaller diameter contribute to the switching at a given temperature as compared to a thinner film. Eventually, the energy barrier that is overcome is the same in the different films. This means that electrical switching may be observable in many antiferromagnets just below or at the onset of exchange bias, which can be taken as a simple measure for the thermal stability and the associated switching energy barrier.

Finally, we come back to the read-out mechanism, which we propose to be either due to SMR or PHE, or both. While the PHE would originate in the MnN layer, the SMR would originate in the Pt layer. We calculate the relative transverse resistivity $\rho_\perp / \rho$ for both cases, where we just use the maximum switching amplitudes $\left| \Delta R_\mathrm{a}\right|$. In the PHE case, the transverse voltage can be written as $U_\perp = \rho_\perp I_\mathrm{MnN} / d_\mathrm{MnN}$. Thus, $(\rho_\perp / \rho_\mathrm{MnN})_\mathrm{PHE} \approx 2 \times 10^{-4}$ for 9\,nm MnN thickness. Accordingly, in the SMR case we have $U_\perp = \rho_\perp I_\mathrm{Pt} / d_\mathrm{Pt}$ and $(\rho_\perp / \rho_\mathrm{Pt})_\mathrm{SMR} \approx 0.9 \times 10^{-4}$. The current branching ratio is approximately $I_\mathrm{Pt} / I_\mathrm{MnN} \approx 7.3$. Both numbers are fairly small compared to our previous experiments on Mn$_2$Au (maximum $\rho_\perp / \rho \approx 70 \times 10^{-4}$) and CuMnAs (maximum $\rho_\perp / \rho \approx 14 \times 10^{-4}$) \cite{Meinert2018, MatallaWagner2019}, where PHE is the only possible read-out mechanism. However, they are similar to the SMR amplitude in Pt / NiO upon rotation in a strong magnetic field (maximum $\rho_\perp / \rho \approx 2 \times 10^{-4}$ at room temperature)\cite{Moriyama2018, Hoogeboom2017}. On the other hand, the SMR can be much larger (maximum $\rho_\perp / \rho \approx 16 \times 10^{-4}$ at room temperature) in YIG / Pt films \cite{Althammer2013}.  Additionally, we performed density functional theory calculations of antiferromagnetic MnN with the fully relativistic multiple-scattering Green function framework as implemented in the SPR-KKR program \cite{Ebert2011, SPRKKR}. We calculated the resistivity tensor via the Kubo-Bastin formalism at finite temperature of 300\,K and determined the PHE amplitude. Lattice vibrations were treated in the alloy analogy model using the coherent potential approximation \cite{Ebert2015, SPRKKR_details}. The mean resistivity was found to be $\mathrm{Tr}(\rho) \approx 57.7\,\mu\Omega\mathrm{cm}$, which is much smaller than the observed value, but still rather high for a metal. The larger resistivity of the thin films arises from the small grain diameter and additional scattering in the grain boundaries. The PHE amplitude is $\rho_\perp / \rho \approx 5.4 \times 10^{-4}$ for $\bm{L} \parallel [100]$ of the face-centered tetragonal unit cell depicted in Fig. \ref{Fig1}\,a). In contrast, with $\bm{L} \parallel [110]$ we obtain $\rho_\perp / \rho \approx -1.7 \times 10^{-4}$. This result indicates that PHE and SMR are of similar magnitude in our system and both may contribute to the signal. However, the theoretical PHE amplitude appears somewhat too small, given that only a small fraction of the film contributes to the observed signal. It is only a factor of 2.5 smaller than the theoretical result, which should be observed when the N\'eel states of all grains are aligned along [100]. This leaves room for speculation whether the anomalous Hall effect due to slightly noncollinear order might contribute to the electrical read-out.  In this case, the small magnetic moment would have to be switched together with the N\'eel order. On the other hand, thermomagnetic effects such as the spin Seebeck effect should not contribute to the signal. This is because we use a lock-in technique and measure the first harmonic signal, whereas thermomagnetic effects would be seen on the dc and second harmonic components. Furthermore, the calculation might underestimate the PHE, because we do not explicitly model chemical disorder nor grain boundary and surface effects. To gain further insight into this open question, a detailed study of the magnetoresistance in strong magnetic fields is necessary and will be performed in the future. 

\section{Summary}
In conclusion, we observe spin-Hall driven electrical switching of the N\'eel order in a metallic, polycrystalline antiferromagnetic layer. The characteristics are fully compatible with a thermal activation model. Our work demonstrates that the characteristic switching properties observed in epitaxial Mn$_2$Au, CuMnAs, or NiO / Pt films can also be obtained in much simpler, polycrystalline antiferromagnetic films. 

\section{Acknowledgements}

\begin{acknowledgments}
We thank the Ministerium f\"ur Innovation, Wissenschaft und Forschung des Landes Nordrhein-Westfalen (MIWF NRW) for financial support. MM acknowledges financial support from the Deutsche Forschungsgemeinschaft (DFG) under sign number ME 4389/2-1. We further thank G. Reiss for fruitful discussions and for making available laboratory equipment. We finally thank H. Ebert for making available the SPR-KKR program.
\end{acknowledgments}


\begin{thebibliography}{30}
\bibitem{SPRKKR_details} The exchange-correlation functional was  the generalized-gradient approximation by Perdew, Burke, and Ernzerhof. A $+U$ term was added to the Mn 3d orbitals with $U = 1.8$\,eV ($J = 0$) in the fully rotationally invariant formulation. This value of $U$ reproduces the experimental lattice geometry and magnetic moments accurately. The Green function was expanded up to $\ell_\mathrm{max} = 3$ and the Fermi energy was accurately obtained with Lloyd's formula. The atomic sphere approximation (ASA) was used throughout. For the evaluation of the Kubo-Bastin formula 32 points were used for the energy integration. Approximately $10^6$ k-points in the full Brillouin zone were used to ensure an accurate evaluation of the Brillouin zone integrals for the Fermi surface term.
\end{thebibliography}
\end{document}